\documentclass[twocolumn,showpacs,preprintnumbers,amsmath,amssymb]{revtex4}

\usepackage{graphicx,psfrag}
\usepackage{amssymb}
\usepackage{amsmath}
\usepackage[squaren,Gray,textstyle]{SIunits}
\usepackage{dcolumn}
\usepackage{bm}

\def\nc{\newcommand}

\def\lsim{\mathrel{\raise.3ex\hbox{$<$\kern-.75em\lower1ex\hbox{$\sim$}}}}
\def\gsim{\mathrel{\raise.3ex\hbox{$>$\kern-.75em\lower1ex\hbox{$\sim$}}}}
\nc{\half}{\frac{1}{2}}
\nc{\shalf}{\ensuremath{\textstyle \frac{1}{2}}}

\nc{\deldag}{\mathbin{\partial\mkern-10.5mu\big/}}
\nc{\kdag}{\mathbin{k\mkern-10mu\big/}}
\nc{\Pdag}{\mathbin{P\mkern-10mu\big/}}

\nc{\beq} {\begin{equation}}
\nc{\eeq} {\end{equation}}
\nc{\beqa}{\begin{eqnarray}}
\nc{\eeqa}{\end{eqnarray}}

\begin{document}


\title{The interior spacetimes of stars in Palatini $f(R)$ gravity}
\author{Kimmo Kainulainen${}^{1}$}
   \email{Kimmo.Kainulainen@phys.jyu.fi}
\author{Vappu Reijonen${}^{2}$}
   \email{Vappu.Reijonen@helsinki.fi}
\author{Daniel Sunhede${}^{1}$}
   \email{Daniel.Sunhede@phys.jyu.fi}
   
\affiliation{${}^{1}$Dept.~of Physics, P.O.~Box 35 (YFL),
   		FIN-40014 University of Jyv\"askyl\"a \\
	${}^{2}$Helsinki Institute of Physics and Dept.~of Physical Sciences,
		P.O.~Box 64, FIN-00014 University of Helsinki, Finland}

\date{\today}

\begin{abstract}

We study the interior spacetimes of stars in the Palatini formalism of $f(R)$ gravity and derive a generalized Tolman-Oppenheimer-Volkoff and mass equation for a static, spherically symmetric star. We show that matching the interior solution with the exterior Schwarzschild-De Sitter solution in general gives a relation between the gravitational mass and the density profile of a star, which is different from the one in General Relativity. These modifications become neglible in models for which $\delta F(R) \equiv \partial f/\partial R - 1$ is a decreasing function of $R$ however. As a result, both Solar System constraints and stellar dynamics are perfectly consistent with $f(R) = R - \mu^4/R$.

\end{abstract}


\maketitle

%
%

\section{Introduction}
\label{sect:Intro}

Current observations indicate that the expansion rate of the universe is accelerating~\cite{astier,spergel}, which in Einstein's theory of General Relativity (GR) implies the existence of a non-zero vacuum energy density, the cosmological constant. Predicting a finite, small vacuum energy is a very serious problem from the quantum field theory point of view, in addition to which the cosmological constant model requires huge amount of fine-tuning to be able to explain the observed acceleration at present. One way to alleviate these  problems is to assume a zero vacuum energy density (due to some yet unknown reason) and instead modify the gravitational action so that dynamics may give rise to the observed acceleration. One popular class of such models consists of generalizing the gravitational action to also contain non-linear interactions in the Ricci scalar $R$:
\begin{equation}
	S = \frac{1}{16 \pi G} \int {\rm d}^{4}x \sqrt{-g} f(R)
	 + \int {\rm d}^{4}x \sqrt{-g} \mathcal{L}_{\rm m} (g_{\mu \nu}) \: ,
\label{eq:action}
\end{equation}
where $\mathcal{L}_{\rm m}$ is the matter Lagrangian. Setting $f(R) = R - 2\Lambda$ corresponds to the canonical Einstein-Hilbert action. It has been shown that the addition of a $1/R$ term to the Einstein-Hilbert action can give rise to the observed acceleration at present~\cite{vollick,carroll}.

In this approach one naturally also has to check if the modified gravity theory is compatible with other observations. Indeed, as was shown by Chiba~\cite{chiba}, using the equivalence between $f(R)$ gravity, understood as a pure {\em metric} theory, and an $\omega = 0$ Brans-Dicke model, a $1/R$ modification to the Einstein-Hilbert action is not compatible with the post-Newtonian (PPN) constraints from Solar System measurements. This conclusion was confirmed, after a recent debate in the literature~\cite{debate,allemandi,ruggiero}, by Erickcek, Smith \& Kamionkowski~\cite{erickcek}, who presented an explicit solution of the field equations arising from the action (\ref{eq:action}) when considered as a pure metric theory. The conclusion of Erickcek \emph{et al.}~has since been shown to also apply for more general forms of $f(R)$~\cite{jin}.

In the metric formalism one assumes that the affine connection of the spacetime manifold is the Levi-Cevita connection, that is, the Christoffel symbol of the second kind $\left\{ {}^{\phantom{.}\rho}_{\mu \nu} \right\}$:
\begin{equation}
	\Gamma^{\rho}_{\mu \nu}
		\equiv \left\{ {}^{\phantom{.}\rho}_{\mu \nu} \right\}
		= \frac{1}{2}g^{\rho \sigma} (\partial_{\mu}g_{\sigma \nu}
			+ \partial_{\nu}g_{\sigma \mu} - \partial_{\sigma}g_{\mu \nu}) \: .
\label{eq:christoffel}
\end{equation}
This choice corresponds to making sure that the metric geodesics, corresponding to the shortest (or strictly speaking extremal) paths between two points on the manifold, coincide with the the affine geodesics, i.e.~the curves for which the tangent is always parallel transported to itself.

However, the restriction (\ref{eq:christoffel}) is by no means obligatory, and more general spacetimes can be studied if the the affine connection is allowed to be a free variable independent of the metric. Not surprisingly, the solutions to the equations of motion arising from the action (\ref{eq:action}) in this larger space of manifolds, are in general different from those obtained in the restricted set of spacetimes corresponding to (\ref{eq:christoffel}). This extension of the configuration space is what is usually called the ``Palatini-formalism'', and it is the point of view to $f(R)$ gravity taken in this paper. We shall refer to this extension of the possible spacetimes as ``metric-affine'' configuration space, in contrast with the ``$g$-metric'' configuration space corresponding to (\ref{eq:christoffel}). For earlier work on gravity in the Palatini formalism, sometimes also referred to as the first order formalism, see~\cite{vollick,palatini,voll_vs_flan}.

Varying the action (\ref{eq:action}) with respect to both the metric $g_{\mu \nu}$ and the affine connection $\Gamma^{\rho}_{\mu \nu}$ results in the equations of motion for Palatini $f(R)$ gravity:
\begin{eqnarray}
	F(R)R_{\mu \nu} - \frac{1}{2}f(R)g_{\mu \nu} & = &
		8\pi G T_{\mu \nu}  \: , 
\label{eq:eomg} \\
	\nabla_{\rho}(\sqrt{-g}F(R)g^{\mu \nu}) & = & 0 \: , 
\label{eq:eomGamma}
\end{eqnarray}
where $F \equiv \partial f/\partial R$, $R \equiv g^{\mu \nu} R_{\mu \nu}(\Gamma)$, $T_{\mu \nu}$ is the energy-momentum tensor of $\mathcal{L}_{\rm m}$, and $\nabla_{\mu}$ is the covariant derivative with respect to the affine connection $\Gamma^{\rho}_{\mu \nu}$. Note that (\ref{eq:eomGamma}) reduces to the $g$-metric compatibility equation if $F(R) = {\rm const}$.  In this case, and only in this case, does the extremal solution for the full metric-affine configuration space happen to reside within the subspace of $g$-metric manifolds. The free affine connection then reduces to the usual Levi-Civita connection (\ref{eq:christoffel}) and the $f(R)$ model collapses to the usual Einstein-Hilbert action, possibly with a cosmological constant.

As was emphasized by Erickcek \emph{et al.}~when studying metric $f(R)$ gravity in Ref.~\cite{erickcek}, a unique exterior solution for a stellar object is found by matching it with an interior solution in the presence of the matter sources. Indeed, the failure to find a metric interior solution that could be matched to an asymptotic Schwarzschild-De Sitter spacetime is the reason why most $f(R)$ theories of gravity are ruled out in the metric formalism~\cite{jin}. Motivated by this observation we will here study the complete spacetimes of spherically symmetric static stars in the Palatini formalism. The hope is of course that new solutions can be found beyond the subspace of $g$-metric manifolds, where the interior solution {\em can} be matched to an exterior metric. This indeed turns out to be the case~\cite{barraco}. 

We will in this paper match the spherically symmetric vacuum solution to the stellar interior in the Newtonian limit. In particular, section II will present an explicit solution for the case of a constant density. In section III, we will derive the exact Tolman-Oppenheimer-Volkoff and mass equation for $f(R)$ gravity in the Palatini formalism, from which the equilibrium configuration of a relativistic star can be computed once the equation of state is given. Unlike the case in GR, the mass equation can not be expressed as a single integral, but is instead cast in the form of a differential equation. We find that in the Palatini formalism, the stellar interior can always be matched to an exterior Schwarzschild-De Sitter solution, whereby these models easily pass the Solar System PPN-constraints~\footnote{Many papers have argued that $f(R)$ gravity in the Palatini formalism passes the Solar System tests without the reference to an explicit interior solution~\cite{allemandi,ruggiero,barraco,solarsystem}. Although matching was considered in the linear approximation in Ref.~\cite{barraco}, this paper did not discuss $1/R$ models, nor did they derive the hydrostatic equlibrium equations for $f(R)$ stars.}. However, we also find that Palatini $f(R)$ models lead to a different relation between the apparent mass of the star and the interior density profile than does General Relativity. As a result, it will in principle be possible to constrain Palatini $f(R)$ models based on astrophysical considerations. However, it turns out that for any model for which $\delta F(R) \equiv F(R)-1$ is a decreasing function of $R$, these corrections are neglible and the stellar interior becomes very similar to the usual solution in GR. In particular, the model with $\delta f(R) \sim 1/R$ is perfectly consistent with both the existing Solar System constraints and the stellar astrophysics.

%
%

\section{Constant density solutions}
\label{sect:solutions}

Let us first consider the solutions to the field equations (\ref{eq:eomg}-\ref{eq:eomGamma}) when the density $\rho$ is constant and pressure $p$ can be neglected. Taking the trace of (\ref{eq:eomg}) one finds a purely algebraic equation relating $R$ and 
$T$:
\begin{equation}
	F(R)R - 2f(R) = 8\pi G T  \: 
\label{eq:trace} \: .
\end{equation}
Obviously, in the case of a constant $T(\rho)$, the Ricci scalar is also a constant and the full field equations (\ref{eq:eomg}-\ref{eq:eomGamma}) are equivalent to the regular Einstein equations with a scaled, $\rho$ dependent gravitational constant $G/F_\rho$:
\begin{equation}
	G_{\mu \nu} + \Lambda_\rho g_{\mu \nu}
		= \frac{8\pi G}{F_\rho} T_{\mu \nu} \: ,
\label{eq:einstein}
\end{equation}
where $G_{\mu \nu} \equiv R_{\mu \nu} - \frac{1}{2}R_\rho g_{\mu \nu}$ and the effective cosmological constant $\Lambda_\rho$ is given by
\begin{equation}
	\Lambda_\rho \equiv \frac{1}{2}\left( R_\rho - \frac{f_\rho}{F_\rho} \right) \: .
\label{lambdarho}
\end{equation}
Since (\ref{eq:eomGamma}) now implies that the affine connection is given by (\ref{eq:christoffel}), the Einstein tensor $G_{\mu \nu}$ can be computed in the usual way from $g_{\mu \nu}$.

We now want to study the general, spherically symmetric solutions for the metric $g_{\mu\nu}$, which can be written in the form:
\begin{equation}
	ds^2 \equiv g_{\mu \nu} x^{\mu} x^{\nu} =
		-e^{A(t,r)}{\rm d}t^2 + e^{B(t,r)}{\rm d}r^2 + r^2{\rm d}\Omega^2 \: .
\label{eq:metric}
\end{equation}
Let us first consider the exterior solution. Because the relation between $R$ and $T$ is purely algebraic, the interior solution will not affect the vacuum value of $R=R_0$. Furthermore, since in vacuum $T_{\mu\nu} = 0$, the problem is reduced to solving the Einstein equations~(\ref{eq:einstein}) in vacuum:
\begin{equation}
	G_{\mu}^{\phantom{\mu}\nu}
		+ \delta_{\mu}^{\phantom{\mu}\nu}\Lambda_0 = 0 \: ,
\label{eq:einsteinvacuum}
\end{equation}
where $\Lambda_0$ is given by (\ref{eq:trace}) and (\ref{lambdarho}) with $T_{\mu\nu}=0$. For example, if $f(R)=R-\mu^4/R$, one finds $\Lambda_0 = \pm \sqrt{3}\mu^2/4$. It is well known that the solution to equation (\ref{eq:einsteinvacuum}) is the static Schwarzschild-dS metric (or Schwarzschild-AdS if $\Lambda_0 < 0$):
\begin{eqnarray}
	e^{A} & = & C(t) \left(
			1 - \frac{2GM}{r} - \frac{\Lambda_0}{3}r^2
		\right) \: , \label{eq:Avacuum} \\
	e^{B} & = & \left(
			1 - \frac{2GM}{r} - \frac{\Lambda_0}{3}r^2
		\right)^{-1} \: . \label{eq:Bvacuum}
\end{eqnarray}
Here the integration constant $M$ is time-independent since $G_{0}^{\phantom{0}1} = 0$ implies that $\partial{B}/\partial t = 0$~\footnote{In $f(R)$ theories of gravity, $\partial{B}/\partial t$ is only zero when $F(R) = {\rm const.}$, so that spherical symmetry alone does not guarantee that the exterior metric is static. In the metric formalism, this signals the breakdown of Birkhoff's theorem, which states that (in GR) a spherically symmetric vacuum solution is necessarily static. However, excluding the conformally invariant case $f(R) \propto R^2$, Birkhoff's theorem still holds in the Palatini formalism, since the trace equation~(\ref{eq:trace}) will fix $R$ to a constant when $T_{\mu \nu} = 0$.}.
Note that introducing the Lagrangian parameter $G$ in (\ref{eq:Avacuum}) and (\ref{eq:Bvacuum}) merely defines a convenient choice of units for $M$.  The other integration constant $C(t)$ is usually absorbed into a redifinition of $t$, but since we want to match the vacuum solution to a stellar interior, we will keep it explicit for the moment.

The gravitational field is sourced by the trace of the energy-momentum tensor $T = -\rho + 3p$ via (\ref{eq:trace}). For all ordinary, non-relativistic stars, $p \ll \rho$ (for example, at the center of the Sun $p \sim 10^{-9} \rho$) and we can hence neglect pressure when solving for the metric. Moreover, we are here considering an idealized object with a constant density, so that $T = \rm {const}$. For this case the Einstein equations are again of the form (\ref{eq:einstein}):
\begin{equation}
	G_{\mu}^{\phantom{\mu}\nu}
		+ \delta_{\mu}^{\phantom{\mu}\nu}\Lambda_\rho
	= - \frac{8\pi G}{F_\rho} \rho u_{\mu}u^{\nu} \: .
\label{eq:einsteinrho}
\end{equation}
It is straightforward to show that the solution to (\ref{eq:einsteinrho}) which is flat and non-singular for an observer at the origin is
\begin{eqnarray}
	e^{A} &=& 
	   \left(
		1 - \frac{2Gm(r)}{r} - \frac{\Lambda_\rho}{3}r^2
	\right)^{-\frac{1}{2} \left(1 - \frac{3\Lambda_\rho}{
		8\pi G\rho/F_\rho + \Lambda_\rho} \right)} , \phantom{apa} \\
\label{eq:Arho}
	e^{B} &=& \left(
		1 - \frac{2Gm(r)}{r} - \frac{\Lambda_\rho}{3}r^2
		       \right)^{-1} ,
\label{eq:Brho}
\end{eqnarray}
where
\begin{equation}
	m(r) \equiv \int_0^r{\rm d}r' \frac{4\pi r'^2 \rho}{F_\rho} \: ,
\label{eq:m}
\end{equation}
and $\Lambda_\rho$ can again be found via (\ref{eq:trace}) and  (\ref{lambdarho}) after the function $f(R)$ is specified. This completely fixes the stellar interior so that the only unknown quantities in the full solution are the parameters $M$ and $C(t)$ of the exterior vacuum. Matching at $r = r_{\odot}$ gives 
\begin{equation}
	M = m(r_{\odot}) + \frac{\Lambda_\rho - \Lambda_0}{6G}r_{\odot}^3 \: ,
\label{eq:M}
\end{equation}
and $C(t) = \left(1 - 2GM/r_{\odot} - \Lambda_{0}r_{\odot}^2/3 \right)^{
			-\frac{3}{2}\big(1 - \frac{\Lambda_{\rho}}{8\pi G \rho/F_{\rho}
				+ \Lambda_{\rho}}\big)}$, i.e. a constant independent of time, which can be absorbed by a simple scaling of $t$ so that the exterior solution is the standard Schwarzschild-dS metric.

The possibility of finding a matching solution is of course obvious in the Palatini formalism. It follows directly from the uniqueness of the exterior solution and the continuous nature of fixing $R$ as a function of $T$, both properties following directly from the trace equation (\ref{eq:trace}). However, what is not trivial is the nonstandard relation between the exterior mass parameter $M$ and the interior density $\rho$, as shown in (\ref{eq:M}) together with (\ref{eq:m}). Since the local density is what defines the local pressure and other thermodynamical properties of the star, it is obvious that $F$ can not differ significantly from $1$ inside the Sun without changing the predictions of Solar physics. Note that although the contribution to $M$ due to the interior effective cosmological constant $\Lambda_{\rho}$ will vanish in the limit $F_{\rho} \rightarrow 1$, the term with $\Lambda_0$ will remain so that $M$ will still differ from the GR value. This is simply due to the fact that $F = 1$ corresponds to GR \emph{without} a cosmological constant, whereas in general the vacuum value of $F = {\rm const.} \neq 1$ so that the theory corresponds to GR with a cosmological constant $\Lambda_0$. Given that $\Lambda_0$ correspond to the observed amount of dark energy $\Omega_{\Lambda} \approx 0.72$~\cite{spergel}, this subtle shift in mass is nevertheless completely neglible.

Interestingly, for all reasonable models for which $\delta F(R) \equiv F(R)-1$ decreases as a function of $R$, $F_\rho$ tends very strongly towards unity. Indeed, for the particular case $f(R) = R - \mu^4/R$, we find from the trace equation~(\ref{eq:trace}) that 
\begin{equation}
R = \frac{1}{2}\left(8\pi G\rho \pm \sqrt{(8\pi G\rho)^2 + 12\mu^4}\right)
\label{eq:R-rho}
\end{equation}
in the Newtonian limit. If the exterior solution is taken to be asymptotically De Sitter space we must pick the solution with a positive sign in (\ref{eq:R-rho}). Hence, given that $\mu^4$ is scaled to cause the observed acceleration, then $\mu^2 \ll 8\pi G\rho$ and hence $F - 1  = \mu^4/R^2 \ll 1$ inside the Sun. In this case the changes to the stellar physics are completely neglible in a $1/R$ model. Purely on academic interest, we note that the situation would be markedly different if one tried to match the $f(R)$ interior solution with an external {\em anti}-De Sitter solution. Here, to match $R$ smoothly to the limiting AdS-value, one would have to take the negative sign in (\ref{eq:R-rho}). Then $R$ would tend to zero and $F$ become very large inside the star. This would imply that the maximum value for the exterior mass parameter of a star with radius $r_\odot$, is constrained to be on the order of $2GM \sim \mu^2 r_\odot^3$.

%
%

\section{The Tolman-Oppenheimer-Volkoff equation}
\label{sect:OV}

Let us now consider the general case of a relativistic star in static equilibrium with an equation of state $p = p(\rho)$. We start with the conservation law for energy-momentum which is of the regular form also in the Palatini formalism of $f(R)$ gravity~\cite{koivisto}. That is, $\widetilde{\nabla}_{\mu} T^{\mu \nu} = 0$, where $\widetilde{\nabla}_{\mu}$ is the covariant derivative with respect to the metric connection $\left\{ {}^{\phantom{.}\rho}_{\mu \nu} \right\}$. Assuming spherical symmetry and that we can describe matter inside the star by a perfect fluid, $T^{\mu \nu} = (\rho + p)u^{\mu}u^{\nu} - pg^{\mu \nu}$, then gives
\begin{equation}
	p' = -\frac{A'}{2}(\rho + p) \: ,
\label{eq:TOVA}
\end{equation}
where a prime denotes differentiation with respect to $r$. In GR one can use the regular Einstein equations to express $A'$ in terms of $\rho$, $p$ and a mass function $m(r)$ similar to (\ref{eq:m}), but with $F_\rho\equiv 1$, after which (\ref{eq:TOVA}) gives the usual Tolman-Oppenheimer-Volkoff (TOV) equation for hydrostatic equilibrium.

In order to find the generalized TOV equation for $f(R)$ gravity, we also need to solve $A'$ from the field equations (\ref{eq:eomg}-\ref{eq:eomGamma}). From (\ref{eq:eomGamma}) it follows that
\begin{eqnarray}
	R_{\mu \nu}(\Gamma) & = & \widetilde{R}_{\mu \nu}(g)
	+ \frac{3}{2}\frac{1}{F^2}
	(\widetilde{\nabla}_{\mu}F)(\widetilde{\nabla}_{\nu}F)
	\nonumber \\
	& & {}- \frac{1}{F} \left(
			\widetilde{\nabla}_{\mu} \widetilde{\nabla}_{\nu}
			- \frac{1}{2}g_{\mu \nu} \widetilde{\Box}
		\right) F \: ,
\label{eq:conformal}
\end{eqnarray}
where a tilde denotes that a quantity is defined in terms of the metric connection $\left\{ {}^{\phantom{.}\rho}_{\mu \nu} \right\}$. We can hence rewrite the field equations (\ref{eq:eomg}) as
\begin{eqnarray}
	\widetilde{G}_{\mu \nu} & = & \frac{8\pi G}{F}T_{\mu \nu}
		- \frac{1}{2}g_{\mu \nu} \left(R - \frac{f}{F} \right) \nonumber \\
	& & {}- \frac{3}{2}\frac{1}{F^2} \left(
			(\widetilde{\nabla}_{\mu}F)(\widetilde{\nabla}_{\nu}F)
			- \frac{1}{2}g_{\mu \nu} (\widetilde{\nabla}F)^2
		\right)
	\nonumber \\
	& & {}+ \frac{1}{F} \left(
			\widetilde{\nabla}_{\mu} \widetilde{\nabla}_{\nu}
			- g_{\mu \nu} \widetilde{\Box}
		\right) F \: ,
\label{eq:eomgmetric}
\end{eqnarray}
where $\widetilde{G}_{\mu \nu} \equiv \widetilde{R}_{\mu \nu} - \frac{1}{2}\widetilde{R}g_{\mu \nu}$. Note that $f$ and $F$ are still functions of the Ricci scalar $R \equiv g^{\mu \nu} R_{\mu \nu}(\Gamma)$, and hence algebraic functions of $T$ via the trace equation~(\ref{eq:trace}).

Using the $00$ and $11$ components of the field equations~(\ref{eq:eomgmetric}) for a static, spherically symmetric metric, one obtains the following source equations for $A(r)$ and $B(r)$:
\begin{eqnarray}
	A' & = & \frac{-1}{1 + \gamma} \left(
			\frac{1 - e^B}{r} - \frac{e^B}{F}8\pi Grp
			+ \frac{\alpha}{r}
		\right) \: , 
\label{eq:sourceA}Ê\\
	B' & = &  \frac{1}{1 + \gamma} \left(
			\frac{1 - e^B}{r} + \frac{e^B}{F}8\pi Gr\rho
			+ \frac{\alpha + \beta}{r}
		\right) \: ,
\label{eq:sourceB}
\end{eqnarray}
where
\begin{eqnarray}
	\alpha & \equiv & r^2 \left(
			\frac{3}{4}\left(\frac{F'}{F}\right)^2  + \frac{2F'}{rF}
			+ \frac{e^B}{2} \left( R - \frac{f}{F} \right)
		\right) \: , 
\label{eq:alpha} \\
	\beta & \equiv & r^2 \left(
			\frac{F''}{F} - \frac{3}{2}\left(\frac{F'}{F}\right)^2
		\right) \: , 
\label{eq:beta}
\end{eqnarray}
and $\gamma \equiv rF'/2F$. 
The TOV equation for $f(R)$ gravity in the Palatini formalism now follows from (\ref{eq:TOVA}) and (\ref{eq:sourceA}):
\begin{eqnarray}
	p' &=& -\frac{1}{1 + \gamma} \phantom{.}
	\frac{(\rho + p)}{r(r - 2Gm_{\rm tot})}
\label{eq:TOV} \\
	& & \times \left( Gm_{\rm tot} + \frac{4\pi Gr^3 p}{F}  
          - \frac{\alpha}{2} (r - 2Gm_{\rm tot} ) 
		 \right) \: , 
\nonumber 
\end{eqnarray}
where we defined a new mass parameter
\begin{equation}
	2Gm_{\rm tot}(r) \equiv r(1 - e^{-B}) \: .
\label{eq:mtot}
\end{equation}
Unlike in GR, it is in general not possible to obtain an explicit solution for $B(r)$, or equivalently, $m_{\rm tot}(r)$ can no longer be found via a simple integration. However, we can use (\ref{eq:sourceB}) to obtain a differential equation for $m_{\rm tot}(r)$:
\begin{eqnarray}
	m_{\rm tot}' & = & \frac{1}{1 + \gamma} \bigg(
		\frac{4\pi r^2\rho}{F} + \frac{\alpha + \beta}{2G}
\label{eq:mass} \\
	& & \phantom{Hanna}
		{}- \frac{m_{tot}}{r}(\alpha + \beta - \gamma) \bigg) \: ,
\nonumber
\end{eqnarray}
where $B$ is completely eliminated in favour of $m_{\rm tot}$ (by eventually using (\ref{eq:mtot}) in the expression (\ref{eq:alpha}) for $\alpha$). Given the equation of state $p(\rho)$, the TOV and mass equation~(\ref{eq:TOV}) and (\ref{eq:mass}) will completely determine $p(r)$, $\rho(r)$, and $m_{\rm tot}(r)$, and hence also the metric via definition (\ref{eq:mtot}) and the source equation (\ref{eq:sourceA}). The exterior mass parameter $M$ is then given by $M = m_{\rm tot}(r_\odot) - \Lambda_0 r_{\odot}^3/6G$, where $\Lambda_0$ is the background, vacuum value of the effective cosmological constant.

\begin{figure}[!t]  
    \begin{center}
	\begin{psfrags}%
	\psfragscanon%
	%
	\psfrag{s03}[t][t]{\setlength{\tabcolsep}{0pt}
		\begin{tabular}{c}$r/2GM_{\odot}$\end{tabular}}%
	\psfrag{s04}[b][b]{\setlength{\tabcolsep}{0pt}
		\begin{tabular}{c}$\lambda \; , \; \rho$\end{tabular}}%

	\psfrag{s03}[t][t]{\setlength{\tabcolsep}{0pt}
		\begin{tabular}{c}$r/2GM_{\odot}$\end{tabular}}%
	\psfrag{s04}[b][b]{\setlength{\tabcolsep}{0pt}
		\begin{tabular}{c}$\lambda \; , \; F \; , \; \rho$\end{tabular}}%
	%
	\psfrag{x01}[t][t]{0}%
	\psfrag{x02}[t][t]{1}%
	\psfrag{x03}[t][t]{2}%
	\psfrag{x04}[t][t]{3}%
	\psfrag{x05}[t][t]{4}%
	\psfrag{x06}[t][t]{5}%
	\psfrag{x07}[t][t]{\shortstack{6\\$\times 10^{6}\ $}}%
	%
	\psfrag{v01}[r][r]{0}%
	\psfrag{v02}[r][r]{0.5}%
	\psfrag{v03}[r][r]{1}%
	\psfrag{v04}[r][r]{1.5}%
	%
	\includegraphics[width=8cm]{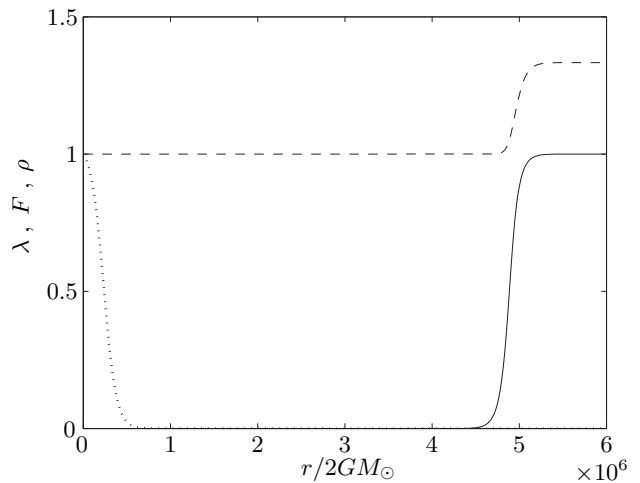}%
	\end{psfrags}%
    \end{center}
    \caption{\small{The effective cosmological constant $\lambda(r)$ (solid)
    	and $F(r)$ (dashed) for $f(R) = R - \mu^4/R$ and an assumed density
    	profile $\rho(r) = \rho_0/(1 + e^{\xi(r - r_{\odot})})$ (dotted),
    	where $\rho_0$ have been chosen so that $M = M_{\odot}$, the radius
		$r_{\odot} = 2.3 \times 10^5$ $(3\textrm{ km}/2GM_{\odot})$, and
		$\xi = 1.4 \times 10^{-5}$.
		Above, $\lambda$ and $\rho$ have been normalized to the observed
		amount of dark energy $\Omega_{\Lambda} \approx 0.72$~\cite{spergel}
		and the central density $\rho(0)$, respectively.}}
    \label{fig:lambda}
\end{figure}
The true equation of state for matter inside a star has to be determined from micro-physical properties, and it is an external input to the gravity sector of the theory. We will not attempt to solve such a problem in this paper. Since pressure is neglible when computing the metric components of an ordinary, non-relativistic star, it is enough for the problem at hand to guess some smooth density profile $\rho(r)$ and then directly use equations (\ref{eq:sourceA}) and (\ref{eq:sourceB}) to determine the metric~\footnote{It is of course possible to assume a density profile also in the case of a relativistic star. However, one then has to use equations (\ref{eq:TOV}) and (\ref{eq:mass}) to first determine the pressure that is needed to support such a configuration.}. For the particular case of $f(R) = R - \mu^4/R$ the trace equation (\ref{eq:trace}) still implies that $F \approx 1$ inside the star, so that the stellar interior should look very much like in GR. One can see how this limit is approached from Fig.~\ref{fig:lambda}, where we plot $F$ and the effective cosmological constant 
\begin{equation}
	  \lambda \equiv \frac{1}{2}\left(R - \frac{f}{F} \right) \,.
\label{eq:lambda}
\end{equation}
as functions of $r$ for a $\rho(r)$ that roughly represents the density profile of a star. Clearly, $F = 1$ inside the star and the contribution to $M$ coming from the effective cosmological constant $\lambda$ will indeed be neglible. Obviously this example confirms the result of the previous section, that a $1/R$ model is completely indistinguishable from GR at the Solar System scale.

%
%

\section{Summary and Discussion}
\label{sect:summary}

We have studied the interior spacetime of stars in the Palatini formalism of $f(R)$ gravity. In particular, we derived the generalized Tolman-Oppenheimer-Volkoff equations for spherical hydrostatic equlibrium. In the Palatini formalism, the trace of the field equations provide an algebraic relation between $R$ and $T$ that gives a unique vacuum solution corresponding to the Schwarzschild-De Sitter metric~\footnote{Excluding the conformally invariant case $f(R) \propto R^2$ for which $R$ remains undetermined.}. Hence, in contrast to metric $f(R)$ gravity, the interior spacetime of a star will not affect the form of the vacuum solution. This is the very reason for why Palatini $f(R)$ gravity can pass the Solar System constraints. However, matching the stellar interior with the exterior vacuum will give a non-standard relation between the gravitational mass and the density of a star. For the particular case $f(R) = R - \mu^4/R$, the effective cosmological constant $\lambda$ vanishes inside the star, i.e. $F \rightarrow 1$, so that the addition of a $1/R$ term to the Einstein-Hilbert action is consistent with Solar System constraints. This is due to the fact that the curvature $R$ becomes large compared to $\mu^2$ inside the star, suppressing the effect of the $\mu^4/R$ term. However, if one considers for example a term $\epsilon R^2$, its effect would grow inside the star. One might thus see interesting modifications to the physics of compact stars, and thereby perhaps be able to put constraints on parameters such as $\epsilon$.

In summary, although one can arrange for different forms of $f(R)$ to give rise to the same cosmological constant in the exterior vacuum solution, the corresponding stellar interior solutions can be vastly different depending on the form of $f(R)$. This raises the possibly to differentiate between different $f(R)$ models and General Relativity, depending on their effect for the stellar evolution and stellar paramers such as luminosity. However, neither Solar System tests nor astrophysical considerations are able to rule out the addition of a $1/R$ term to the Einstein-Hilbert action for gravity.

%
%

\begin{acknowledgments}
D. Sunhede would like to thank several participants of the Les Houches Summer School, session 86, for useful discussions.
This work was partially supported by a grant from the
Magnus Ehrnrooth Foundation (VR).
\end{acknowledgments}

%
%

\end{document}